\documentclass[letterpaper, 10 pt, conference]{ieeeconf}

\IEEEoverridecommandlockouts   

\usepackage{caption}
\usepackage{subcaption}
\usepackage[utf8]{inputenc} 
\usepackage[T1]{fontenc}    
\usepackage{url}            
\usepackage{booktabs}       
\usepackage{amsfonts}       
\usepackage{graphicx}%
\usepackage{amsmath}
\usepackage{comment}
\usepackage{amssymb}
\usepackage{xcolor}
\usepackage{accents}
\usepackage{caption}
\usepackage{cite}
\usepackage{cite}
\usepackage{multirow}
\usepackage{siunitx}
\usepackage{bm}
\usepackage{algorithm}
\usepackage{algpseudocode}
\usepackage{multirow}
\usepackage{footmisc}
\let\labelindent\relax
\usepackage{enumitem}

\usepackage[commandnameprefix=always]{changes}

\newtheorem{theorem}{Theorem}
\newtheorem{problem}{Problem}

\newtheorem{corollary}{Corollary}

\newtheorem{remark}{\textbf{Remark}}

\newtheorem{assumption}{Assumption}

\renewcommand{\baselinestretch}{.99}

\def\qed{ \rule{.1in}{.1in}}

\usepackage{changes}

\title{\LARGE \bf
Distributed Resource Allocation for Human-Autonomy Teaming under Coupled Constraints
}
\author{
\thanks{{\tt\small}.}
}
\author{
Yichen Yao, Ryan Mbagna Nanko, Yue Wang, and Xuan Wang
\thanks{Y. Yao and X. Wang are with the Electrical and Computer Engineering Department, George Mason University; R. Nanko and Y. Wang are with the Mechanical Engineering Department at Clemson University. This work is partially supported by the NSF under EES grant no. 2005030.}%
}

\begin{document}
\maketitle
\thispagestyle{empty}
\pagestyle{empty}
\raggedbottom

\begin{abstract}
This paper studies the optimal resource allocation problem within a multi-agent network composed of both autonomous agents and humans.
The main challenge lies in the globally coupled constraints that link the decisions of autonomous agents with those of humans. To address this, we propose a novel reformulation that transforms these coupled constraints into decoupled local constraints defined over the system’s communication graph. Building on this reformulation, and incorporating a human response model that captures human-robot interactions while accounting for individual preferences and biases, we develop a fully distributed algorithm. This algorithm guides the states of the autonomous agents to equilibrium points which, when combined with the human responses, yield a globally optimal resource allocation.
We provide both theoretical analysis and numerical simulations to validate the effectiveness of the proposed approach.
\end{abstract}


\section{Introduction} 
\begin{figure}[t]
    \centering
    \includegraphics[width=0.7\linewidth]{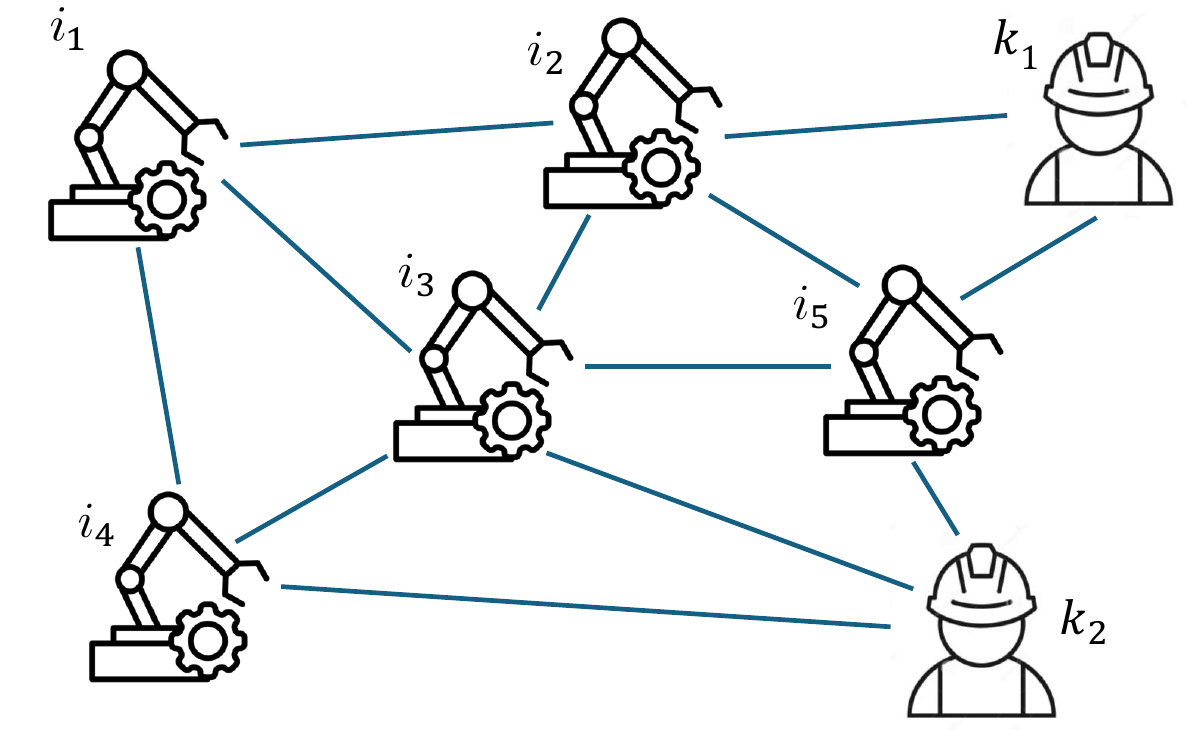}
    \caption{Distributed resource allocation for human-autonomy teaming and the interaction/communication topology.}
    \label{fig:network-structure}
    \vspace{-1.5em}
\end{figure}

Human-autonomy teaming offers significant advantages in real-world systems by leveraging the high precision and efficiency of autonomous agents alongside the intelligence and adaptability of humans~\cite{cao2021human}. A representative application is the multi-agent resource allocation for manufacturing, where autonomous agents and humans coexist and interact through an underlying communication network, as illustrated in Fig.\ref{fig:network-structure}.
In fully autonomous systems, numerous distributed algorithms have been developed to achieve scalable and efficient resource allocation. These methods typically rely on carefully designed update rules for each agent’s state. Because of this, they are not directly applicable when humans are involved, primarily due to the lack of controllability in human behavior and the presence of potential biases in human decision-making~\cite{kwon2020humans}.
In this context, the primary challenge arises from the globally coupled constraints~\cite{wang2018distributed, zhu2011distributed}, which require all autonomous agents to adapt to human behaviors using only local information.
In this paper, we present a new reformulation that can transform 
the globally coupled constraint into a set of decoupled local constraints defined over the system’s communication graph. This reformulation simplifies the integration of human agents (modeled via a response function that captures biases and risk preferences), and enables the use of distributed optimization techniques to achieve resource allocation among autonomous agents and human across the entire network.

\textbf{\textit{Literature review:}} Resource allocation is a fundamental problem in multi-agent systems, with wide-ranging applications in areas such as communication networks~\cite{zulhasnine2010efficient}, energy management~\cite{tang2015resource}, and automated manufacturing~\cite{morariu2020machine}. A key challenge in developing efficient allocation algorithms lies in globally coupled resource constraints across all agents when optimizing their collective behaviors~\cite{zhu2011distributed}.
Traditional centralized methods face scalability issues as the number of agents increases. To address this, dual decomposition methods~\cite{xiao2004simultaneous } allow for local gradient computations and parameter updates, thus reducing the computational load on the centralized coordinator. Nevertheless, because the update of dual variables still depends on the coupled resource allocations of all agents, the centralized coordinator cannot be entirely removed.

To overcome this limitation, many efforts have focused on designing fully distributed algorithms for resource allocation. Most approaches adopt a distributed treatment of the dual variable, i.e., creating local copies of the dual variable, and then introducing coordination mechanisms to ensure their consistency across all agents. Examples include Consensus-ADMM (Alternating Direction Method of Multipliers)~\cite{shorinwa2023distributed}, D-ADMM~\cite{jiang2021distributed} or Tracking-ADMM~\cite{falsone2020tracking}. These approaches have been further extended to handle more complex constraints, including non-identity mappings between agent states and resources~\cite{banjac2019decentralized}. They also address both equality~\cite{nedic2018improved} and inequality constraints~\cite{aybat2019distributed}, local constraints~\cite{zhu2019distributed}, and offer various convergence rates, including sub-linear~\cite{doan2017distributed} and linear~\cite{jiang2021distributed}.
While the distributed implementation of ADMM-based algorithms is theoretically elegant and effective for fully autonomous systems, these methods are not directly applicable when human agents are involved. This is because they typically rely on coordinated primal-dual updates to enforce global constraints. However, human behaviors cannot be explicitly programmed, and instead depend on their subjective preference and bias~\cite{tverskyJudgmentUncertaintyHeuristics1974}.

When humans are involved, instead of designing their behaviors, we can only model them.
Literatures in behavioral economics and psychology offer insights into modeling the intricacies in human decision-making patterns~\cite{ kahnemanThinkingFastSlow2011}, which allows the quantification of human bias and risk attitudes (risk-seeking vs. risk-averse). Expected utility theory has been long used to choose the option with the highest expected payoff based on objective utility measures. However, this neglects human bias and emotion in decision-making and often contradicts observed human behaviors. ~\cite{machinaEncyclopediaActuarialScience2004}.
In contrast, subjective expected utility theory (SEU), prospect theory, regret theory, and decision field theory have been developed to explain and predict human bias and risk attitudes in decision-making~\cite{kobberlingPreferenceFoundationsNonexpected2003}. 
Recent developments have also extended the above theories to computational models using machine learning approaches with engineering applications such as shared mobility, smart grids, and human-autonomy collaboration~\cite{guan_cdc_2019,jiang2022risk,jiang2022risk-TITS}.
As we explore multi-human-multi-robot teaming problems, it is essential to address multialternative decision-making for diverse sets of needs~\cite{roeMultialternativeDecisionField2001a}. However, existing works do not provide a closed-form solution, which is incompatible with most engineering and machine learning algorithms~\cite{hamalainen2023differentiable}.

\textbf{\textit{Statement of contribution:}} 
We study the multi-agent resource allocation problem involving both autonomous agents and humans. Given that human behaviors cannot be engineered, we introduce a human model in the form of a response function to capture their interactions with autonomous agents. This modeling process accounts for human biases and risk attitudes. The primary contribution of this work lies in a reformulation of the globally coupled constraint, which transforms it into a decoupled constraint defined over the system's communication graph. Compared to the distributed treatment of dual variables, this reformulation enables a more straightforward and natural integration of human agents in algorithm design. Based on this reformulation, we develop a fully distributed algorithm for autonomous agents, whose effectiveness is validated both theoretically and through simulated experiments.

\smallskip
\textit{Notations}:
Throughout this paper, we let $\mathbb{R}^n$ denote the $n$-dimensional real vector space; $\mathbb{R}^n_{\ge0}$ means the corresponding vectors have non-negative entries; let $\mathbf{1}_r$ denote the vector in $\mathbb{R}^r$ with all entries equal to 1; let $I_r$ denote the $r \times r$ identity matrix. We let $\text{col}\{A_1, A_2, \cdots, A_r\} = \left.\left[\begin{array}{cccc}{A_{1}^{\top}}&{A_{2}^{\top}}&{\cdots}&{A_{r}^{\top}}\end{array}\right.\right]^{\top}$ denote a stack of matrices $A_i$. We let $\text{diag}\{A_1, A_2, \dots, A_r\}$ denote the block diagonal matrix with $A_i$ as the $i$th diagonal block entry, for $i = 1, 2, \dots, r$. Let $\otimes$ denote the Kronecker product. Let $\| \cdot \|$ denote the norm of a vector or a matrix. 
For scalars $a,b\in\mathbb{R}$, the operator $[a]_b^+$ is defined as:
$$[a]_b^+=
\begin{cases}
a,&\text{if} ~b>0\\
\max\{0,a\}, &\text{if} ~b=0
\end{cases}$$
For vectors $a, b\in\mathbb{R}^n$, $[a]_b^+$
denotes the vector whose $i$-th component is $[a_i]_{b_i}^+$, $i\in\{1,\cdots, n\}.$
For a vector $a\in\mathbb{R}$, the notion $a\le0$ means all the entries of $a$ are non-positive. 

\section{Problem Formulation and Running Example}

As shown in Fig. \ref{fig:network-structure}, consider a system comprising a set $\mathcal{M}=\{i_1,\cdots,i_m\}$ of autonomous agents and a set $\mathcal{H}=\{k_1,\cdots,k_h\}$ of human agents with $|\mathcal{M}|=m$ and $|\mathcal{H}|=h$. We represent the communication and interaction structure among agents using a graph $\mathbb{G}$, as illustrated in Fig. 1. 
The graph $\mathbb{G}$ is assumed to be connected and undirected, with vertex set $\mathcal{V}(\mathbb{G}) = \mathcal{M} \cup \mathcal{H}$. The edge set $\mathcal{E}(\mathbb{G})$ defines neighboring relations, where an edge between any two agents indicates they can observe each other's states and interact.
Let $\mathcal{N}_i$ and $\mathcal{N}_k$ denote the neighbor set of autonomous agent $i\in\mathcal{M}$ and human agent $k\in\mathcal{H}$, respectively. 
Suppose each autonomous agent $i$ possesses a state $x_i\in\mathbb{R}^{n_i}$ and a cost function $f_i(\cdot): \mathbb{R}^{n_i} \to \mathbb{R}$. Each human agent $k$ possesses a state $y_k\in\mathbb{R}^{s_k}$ and a cost function $g_k(\cdot): \mathbb{R}^{s_k} \to \mathbb{R}$. Note that the states of different agents do not necessarily need to have the same dimension. 

The system performance is evaluated by the sum of individual costs incurred by both autonomous and human agents, given by:
\begin{align}\label{eq_obj1}
\sum_{i\in\mathcal{M}} f_i(x_i) + \sum_{k\in\mathcal{H}} g_k(y_k)
\end{align}
In addition, the states $x_i,y_k$ are subject to constraints (e.g., system resources), which are globally coupled among all agents. For simplicity, the coupling of the constraints are assumed to be linear form:
\begin{align}\label{eq_cst1}
\sum_{i\in\mathcal{M}} A_i x_i + \sum_{k\in\mathcal{H}} B_k y_k +c \le 0
\end{align}
where $A_i\in\mathbb{R}^{r\times n_i}$ and $B_k\in\mathbb{R}^{r\times s_k}$ are matrices mapping agents' local states to the overall system constraint $c\in\mathbb{R}^{r}$.

To capture the interaction between humans and autonomous agents, we introduce the following human response model:
\begin{equation}\label{eq_response}
    y_k = q_k(x_{\mathcal{N}_{k}}),
\end{equation}
where the human behaviors $y_k$, introduced in (\ref{eq_obj1}) and (\ref{eq_cst1}), depend on the states of its neighboring autonomous agents.
A detailed justification of this response model will be provided in the next section.


By combining the formulation~(\ref{eq_obj1}–\ref{eq_cst1}) with the human response model~\eqref{eq_response}, our goal is to develop a distributed algorithm that coordinates the collective behavior of all autonomous agents to optimize the overall cost of the human-robot system. This leads to the following formulation.
\begin{problem}\label{prob_1}
Develop a fully distributed algorithm that allows autonomous agents in the system to collaboratively respond to humans and find the optimal states that solve:
\begin{subequations}\label{eq_pf}
\begin{align}
\min_{x_i} \quad &    \sum_{i\in\mathcal{M}} f_i(x_i) + \sum_{k\in\mathcal{H}} g_k(y_k) \label{eq_pfa}\\
    \text{s.t.} \quad &\sum_{i\in\mathcal{M}} A_i x_i + \sum_{k\in\mathcal{H}} B_k y_k + c \le 0 \label{eq_pfb}\\
    & y_k = q_k(x_{\mathcal{N}_{k}}),~~\text{for all $k\in\mathcal{H}$.}\label{eq_pfc}
\end{align}
\end{subequations}
\end{problem}
\smallskip
With human response function~\eqref{eq_pfc}, we assume the Slater’s conditions~\cite{slater2013lagrange} is satisfied for~\eqref{eq_pfb}, that is, there exists $x_i\in \mathbb{R}^{n_i}$ such that $\sum_{i\in\mathcal{M}} A_i x_i + \sum_{k\in\mathcal{H}} B_k q_k(x_{\mathcal{N}_{k}}) + c < 0$.




\textbf{\textit{Running example:}} To conceptualize problem \eqref{eq_pf} within an engineering context, consider a complex manufacturing system. Let $x_i$ and $y_k$ represent the quantities of different products manufactured by autonomous and human-operated nodes, respectively. 
The functions $f_i(\cdot)$ and $g_k(\cdot)$ represent the corresponding manufacturing costs. The matrices $A_i$ and $B_k$ map the production of each node to overall system constraints and demands, which may include total funds, labor/time/transportation resources, productivity requirements, and other relevant factors. 
In this system, humans are uncontrollable, meaning their states, $y_k$, cannot be engineered and instead depend subjectively on their response functions $q_k(\cdot)$. 
Therefore, autonomous nodes must account for human responses and collaboratively design updating strategies for $x_i$ that optimize production strategies, minimize costs and  satisfies these constraints. 



\section{Human Modeling and Virtual Human Proxy}\label{Sec_HM}

In this section, we describe our human modeling principle and methodology. The goal is to model and predict human response in terms of workload or resources allocation given the network structure (see Fig. \ref{fig:network-structure}) and the states of the neighboring autonomous agents. 
In our system, human states $y_k$ are not controllable by algorithms. Instead, we consider it to depend subjectively on their response functions \eqref{eq_response}.

\begin{figure}[t]
    \centering
    \includegraphics[width=0.75\linewidth]{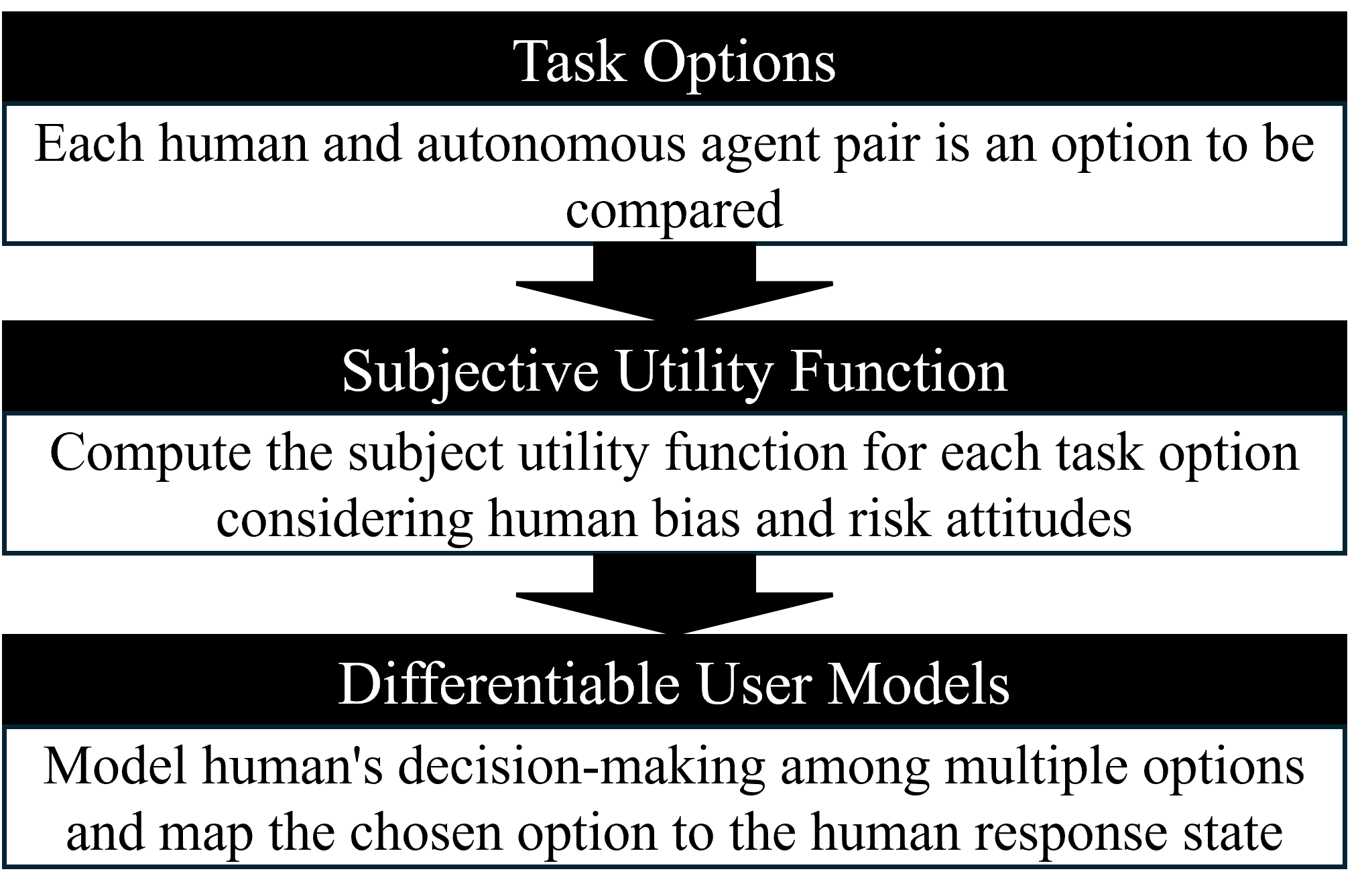}
    \caption{Flowchart of the human modeling process.}
     \vspace{-1.5em}
    \label{fig:HRI_model}
\end{figure}

The overall modeling process to obtain \eqref{eq_response} is summarized as follows\footnote{The exact computational model would require the design of human subject tests subject to IRB approval, data collection, and the development of suitable machine learning approaches to learn the model. The detailed human modeling process will be addressed separately in our future work.} and is shown in Fig.~\ref{fig:HRI_model}. First, we will evaluate the human's subjective utility for each human–autonomous agent pair. Consider a human–autonomy collaborative manufacturing scenario where humans and autonomous agents work together to produce parts~\cite{sadrfaridpour2017collaborative}. Each human and autonomous agent pair can have different workload allocation, depending on tasks the human would like to perform independently and the capabilities of different autonomous agents. To account for human biases and risk attitudes in decision-making, a subjective utility for each human and autonomous agent pair can be shaped by decision-making theories such as prospect theory~\cite{kahnemanProspectTheoryAnalysis1979} and regret theory~\cite{bleichrodt2010quantitative}.

After constructing the subjective utility for each human and autonomous agent pair, we need to model human's decision-making among multiple options and map the chosen option to the human response state $y_k$. There are several different approaches to model and predict such human behaviors in the literature, such as the multi-alternative decision field theory \cite{busemeyerDecisionFieldTheory1993, roeMultialternativeDecisionField2001a}, and regret theory with general choice sets\cite{quiggin1994regret}. However, these models are incompatible with most machine learning and engineering algorithms in practice~\cite{hamalainen2023differentiable}. Therefore, in order to find the optimal allocation solution for human-autonomy teaming, we will assume human response functions that are differentiable. A suitable approach is a differentiable neural network function with human subject utility as the input and human desired workload as the output. The integration of the subjective utility in human decision-making with the differentiability of black-box artificial neural networks creates a trade-off between model explainability and computational feasibility. 

In terms of algorithm design and implementation for autonomous agents, due to the fact that human states $y_k$ need to be shared with connected autonomous agents in network $\mathbb{G}$, we introduce a virtual human proxy for each human agent. These virtual proxies approximate human response functions $q_k(\cdot)$ as introduced above, facilitate information sharing, and will support some auxiliary state (other than $y_k$) updates. Note that these virtual human proxies do not need to physically exist. They can be assigned to one of the human's neighboring autonomous agents, if it can access the same information as the human agent.

\begin{remark}[Human Response Functions]\label{RM_H}
In this paper, we assume human response functions $q_k(x_{\mathcal{N}_{k}})$ are known. However, in practice, the virtual human proxy might need a learning process before obtaining the true $q_k(x_{\mathcal{N}_{k}})$. 
To address this, we will introduce a Corollary \ref{col_1} to demonstrate that the proposed algorithm remains applicable, if it is implemented with an approximation of $q_k(x_{\mathcal{N}_{k}})$ that is inaccurate at the beginning then gradually converge to its true value. 

As another remark, our response function $y_k = q_k(x_{\mathcal{N}_{k}})$ in \eqref{eq_response} considers only the human response depending on neighboring autonomous agents, rather than on the states of other humans, i.e., $y_k = q_k(x_{\mathcal{N}_{k}},y_{\mathcal{N}_{k}})$.
This is due to the intrinsic complexity of bi-directional human-human interaction, which requires game-theoretic modeling to characterize the outcome. However, depending on the form of $q_k(\cdot)$, the Nash equilibrium of the game usually has a closed-form solution purely determined by the behaviors of autonomous agents~\cite{owen2013game}, which simplifies to our formulation \eqref{eq_response}. An explanation of this fact using compact state representation is provided later in Footnote \ref{fnlabel}.
\hfill $\bullet$
\end{remark}

\section{Main Results}
To solve problem \ref{prob_1}, the main challenge lies in the globally coupled constraint \eqref{eq_pfb} defined among all autonomous agents and humans. To address this, we will first introduce an equivalent reformulation that converts the coupled constraint into a decoupled constraint defined over the graph $\mathbb{G}$. Following this, we present a fully distributed algorithm for solving the problem described in \eqref{eq_pf}.
For clarity of presentation, this section focuses solely on the results. All proofs are provided in the following section. We start with an assumption.

\begin{assumption}\label{Asmp_1}
The following conditions hold:
\begin{enumerate}[label={\alph*}.]
    \item The graph $\mathbb{G}$ is undirected and connected.
    \item For all $i\in\mathcal{M}$, $f_i(x_i)$ is differentiable and convex in $x_i$.
    \item For all $k\in\mathcal{H}$, $q_k(x_{\mathcal{N}_k})$ is differentiable and convex in $x_j$, $\forall j\in {\mathcal{N}_k}$.
    \item For all $k\in\mathcal{H}$, $g_k(q_k(x_{\mathcal{N}_k}))$ and $B_k\cdot q_k(x_{\mathcal{N}_k})$ are differentiable and convex in $x_j$, $\forall j\in {\mathcal{N}_k}$.
\end{enumerate}    
\end{assumption}


\textit{Justification}:
Assumptions \ref{Asmp_1}-(a-c) are regular assumptions that have been widely used in the literature for distributed resource allocation \cite{nedic2009distributed,shorinwa2023distributed} and optimization \cite{bertsekasnonlinear}.
Our assumption \ref{Asmp_1}-(d) arises from the consideration of the human response function $q_k(\cdot)$, which leads to a function composition when calculating system cost $g_k(\cdot)$.
To ensure the convexity of the composition, a sufficient condition~\cite[Sec. 3.2.4]{boyd2004convex} is that both $q_k(\cdot)$ and $g_k(\cdot)$ are convex functions and, additionally, $g_k(\cdot)$ is non-decreasing in its arguments. This condition is generally true for a wide class of allocation problems with the goal of cost-minimization, where the cost grows with the agent's states, e.g., quantities of manufactured products in the running example. The same applies to constraint function $B_k\cdot q_k(x_{\mathcal{N}_k})$. 



\subsection{Decoupling the Global Constraint}
For the simplicity of analyzing problem \eqref{eq_pf}, we first rewrite it into a compact form: 
\begin{subequations}\label{eq_repf}
\begin{align}
\min_{\mathbf{x}} \quad & F(\mathbf{x}) + G(\mathbf{y}) \label{eq_repfa} \\
\text{s.t.} \quad & \mathbf{A}\mathbf{x} + \mathbf{B}\mathbf{y} + \mathbf{c} \le 0 \label{eq_repfb} \\
& \mathbf{y} = Q(\mathbf{x}) \label{eq_repfc}
\end{align}
\end{subequations}
Here,
\begin{align*}
    &F(\mathbf{x})=\sum_{i\in\mathcal{M}} f_i(x_i), \quad G(\mathbf{y})=\sum_{k\in\mathcal{H}} g_k(y_k),
\end{align*}
and $\mathbf{x}=\text{col}\{x_{i_1},\cdots,x_{i_m}\}$, $\mathbf{y}=\text{col}\{y_{k_1},\cdots,y_{k_h}\}$ are compact states;
 $\mathbf{A}=\begin{bmatrix}
     A_{i_1}&\cdots&A_{i_m}
 \end{bmatrix}$, $\mathbf{B}=\begin{bmatrix}
     B_{k_1}&\cdots&B_{k_h}
 \end{bmatrix}$ are compact matrices. Also, we have $\mathbf{y}=Q(\mathbf{x})=\text{col}\{q_{k_1}(\cdot), \cdots,q_{k_h}(\cdot)\}$\footnote{Echoing Remark \ref{RM_H}, if we consider human-human interaction, then the human response function should be written as an implicit function $\mathbf{y}=\widehat{Q}(\mathbf{x},\mathbf{y})$. If this equation is solvable with a unique closed-form equilibrium $\mathbf{y}^*=Q(\mathbf{x}^*)$, then this equilibrium can be used in \eqref{eq_repfc}.\label{fnlabel}}.

Equation \eqref{eq_repfb} is globally coupled, where the satisfaction of the constraints requires checking the states for all other agents in the system.
To circumvent this, we present a novel reformulation that transforms \eqref{eq_repfb} into a decoupled form. 

\begin{theorem}
    (\textit{Decoupling the globally coupled constraint over the network}) \label{Thm_0}
Suppose Assumption \ref{Asmp_1}-a holds. The
coupled constrain in \eqref{eq_repfb}, holds if
and only if there exists a vector $\mathbf{z}\in\mathbb{R}^{r(m+h)}$ such that
\begin{align}\label{eq_lm1}
\begin{bmatrix}
        \bar{A}&\\&\bar{B}
    \end{bmatrix}\begin{bmatrix}
        \mathbf{x}\\\mathbf{y}
    \end{bmatrix} +\bar{L}\mathbf{z}+C\le0,
\end{align}
where $\bar{A} = \mathrm{diag}\{A_{i_1},\cdots,A_{i_m}\}$; $\bar{B} = \mathrm{diag}\{B_{k_1},\cdots,B_{k_h}\}$; 
$\bar{L}= L\otimes I_{r}$ with $L\in\mathbb{R}^{(m+h)\times(m+h)}$ being the Laplacian matrix of graph $\mathbb{G}$;
$\mathbf{z}=\text{col}\{z_{i_1}, \cdots, z_{i_m}, {z}^H_{k_1},\cdots,{z}^H_{k_h}\}$ is a compact variable with the first $m$ components $z_i$ correspond to autonomous agents, and the later $h$ components ${z}^H_k$ correspond to human agents.
Using the same notations principle, define $C = \text{col}\{c_{i_1}, \cdots, c_{i_m}, {c}^H_{k_1},\cdots,{c}^H_{k_h}\}$ as any vector satisfying $c_i\in\mathbb{R}^r,{c}^H_k\in\mathbb{R}^r$ and the following condition:
$$\sum_{i\in\mathcal{M}} c_i+\sum_{k\in\mathcal{H}}{c}^H_k  = c.$$ 
An easy way to construct $C$ is to let $c_{i_1}=c$; $c_{i}=0, \forall i\neq i_1$; and ${c}^H_{k}=0, \forall k$. 
\end{theorem} 

\smallskip
Proof of this result will be given in the next section.

\begin{remark}[Decoupled Constraints over Graph]
The reformulation \eqref{eq_lm1} represents a decoupled constraint that will allow a more straightforward and natural integration of human agents for distributed algorithm design.
First, the constraint is defined solely by the local interactions between agents through the network graph. For any autonomous agent $i \in \mathcal{M}$, the constraint is equivalent to:
\begin{align}\label{eq_const_x}
  A_ix_i +\!\!\!\!\sum_{j\in\left(\mathcal{N}_i\bigcap \mathcal{M}\right)}\!\!(z_i-z_j)+
    \!\!\!\!\sum_{\ell \in \left(\mathcal{N}_i\bigcap \mathcal{H}\right)}\!\!(z_i-z^H_\ell)+ c_i \le 0;  
\end{align}
and for any human agent $k \in \mathcal{H}$, the constraint is:
\begin{align}\label{eq_const_y}
  B_k y_k &+ \sum_{j\in\left(\mathcal{N}_k\bigcap \mathcal{M}\right)}(z^H_k-z_j)\nonumber\\
    & +\sum_{\ell \in \left(\mathcal{N}_k\bigcap \mathcal{H}\right)}(z^H_k-z^H_\ell)+ c^H_k \le 0 
\end{align}
For both expressions, each agent's constraint only involves its own state and the states of its neighbors ($j$'s represent the autonomous agent neighbors; $\ell$'s represent human agent neighbors).  This means the global constraint is decoupled over the network. 

To develop an algorithm, for autonomous agents, we will design dynamics for decision variables $x_i$ and auxiliary variables $z_i$ to satisfy the local constraint~\eqref{eq_const_x}. For human agents, the states $y_k$ will follow human models, and the satisfaction of the local constraint~\eqref{eq_const_y} can be ensured by updating only the auxiliary variables $z_k^H$.
\hfill $\bullet$
\end{remark}

\subsection{The Proposed Fully Distributed Algorithm}
Based on the constraint reformulation in Theorem \ref{Thm_0}, solving the problem becomes much simpler, as it is now equivalent to minimizing \eqref{eq_repfa} subject to the decoupled local constraints \eqref{eq_lm1}. To this end, we can employ a \textit{saddle point flow} method that is widely used for distributed optimization and adapt it to consider human agents.
Consider the following Lagrangian function:
\begin{align}\label{eq_Larg}
\mathcal{L}(\mathbf{x},\mathbf{z},\boldsymbol{\lambda})\!=\!
F(\mathbf{x})\!+\!G(Q(\mathbf{x}))\!+\!\boldsymbol{\lambda}^{\mathsf{T}}\left(\begin{bmatrix}\bar{A}\mathbf{x}\\\bar{B}Q(\mathbf{x})\end{bmatrix}\!+\!\bar L\mathbf{z}\!+\!C\right).
\end{align}
Here, $\mathbf{y}$ has been replaced by $Q(\mathbf{x})$ using \eqref{eq_repfc}.
The $\boldsymbol{\lambda}$ is the Lagrange multipliers, defined as $\boldsymbol{\lambda} = \text{col}\{\lambda_1, \cdots, \lambda_{m}, {\lambda}^H_1,\cdots,{\lambda}^H_h\}$ with the first $m$ components $\lambda_i$ corresponding to autonomous agents, and the later $h$ components $\hat{\lambda}_k$ corresponding to human agents. Incorporating these Lagrange multipliers, the constrained optimization problem is transformed into an unconstrained one, and the optimal solution of the original problem is now associated with the saddle point of the Lagrangian function.
To achieve the saddle point of \eqref{eq_Larg}, consider the following flow:

\begin{subequations}\label{eq_alg}
    \begin{align}\label{eq_alga}
    \dot{\mathbf{x}}&=-\frac{\partial\mathcal{L}(\mathbf{x},\mathbf{z},\boldsymbol{\lambda})}{\partial\mathbf{x}}=-\left(\frac{\partial F}{\partial\mathbf{x}}+\frac{\partial Q}{\partial \mathbf{x}}\frac{\partial G}{\partial Q}+
    \begin{bmatrix}
         \bar{A}\\\bar{B}\frac{\partial Q}{\partial \mathbf{x}}
    \end{bmatrix}^{\top}\boldsymbol{\lambda}\right)\\\label{eq_algb}
    \dot{\mathbf{z}}&=-\frac{\partial\mathcal{L}(\mathbf{x},\mathbf{z},\boldsymbol{\lambda})}{\partial \mathbf{z}}=-\bar L\boldsymbol{\lambda}\\\label{eq_algc}
    \dot{\bm{\lambda}}&=\left[+\frac{\partial\mathcal{L}(\mathbf{x},\mathbf{z},\boldsymbol{\lambda})}{\partial\boldsymbol{\lambda}}\right]_{\boldsymbol{\lambda}}^{+}= \left[
    \begin{bmatrix}
        \bar{A}\mathbf{x}\\\bar{B}Q(\mathbf{x}) 
    \end{bmatrix}
 +\bar L\mathbf{z}
     +C\right]_{\boldsymbol{\lambda}}^{+},
\end{align}
\end{subequations}
where $[~\cdot~]_{\boldsymbol{\lambda}}^{+}$ is a projection defined in the \textit{Notations} section. It ensures $\boldsymbol{\lambda}$ is always non-negative. 
Note that the dynamics for $\bm{\lambda}$ is discontinuous; to establish the existence of solutions to this differential equation, one can consider solutions in the Caratheodory sense. A rigorous treatment of this can be found in~\cite{cherukuri2016asymptotic}.

We also provide the agent-based implementation of \eqref{eq_alg} to demonstrate that the algorithm is fully distributed, i.e., the state update of each agent only relies on the information of its own and that of its neighbors. 
Specifically, for any autonomous agent $i\in\mathcal{M}$, one has:
\begin{subequations}\label{eq_algda}
\begin{align}\label{eq_algda1}
   \dot{{x}}_i&=-\frac{\partial f_i}{\partial x_i}-A_i^{\top}\lambda_i \nonumber\\&-\sum_{\ell \in \left(\mathcal{N}_i\bigcap \mathcal{H}\right)} \left( \frac{\partial q_\ell}{\partial x_i}\frac{\partial g}{\partial q_\ell}  + \left(B_\ell \frac{\partial q_\ell}{\partial x_i}\right)^{\top} {\lambda}^H_\ell \right)\\[.5ex]\nonumber\\
    \dot{z}_i&=-\sum_{j\in\left(\mathcal{N}_i\bigcap \mathcal{M}\right)}(\lambda_i-\lambda_j)-
    \sum_{\ell \in \left(\mathcal{N}_i\bigcap \mathcal{H}\right)}(\lambda_i-\lambda^H_\ell) \\[.5ex]\nonumber\\
   \dot{\lambda}_i&= \!\!\left[\!A_ix_i +\!\!\!\!\!\!\sum_{j\in\left(\mathcal{N}_i\bigcap \mathcal{M}\right)}\!\!(z_i-z_j)+
    \!\!\!\!\!\!\sum_{\ell \in \left(\mathcal{N}_i\bigcap \mathcal{H}\right)}\!\!(z_i-z^H_\ell)+ c_i\!\right]_{\lambda_i}^+
\end{align}
\end{subequations}
For any human proxy agent $k\in\mathcal{H}$, one has:
\begin{subequations}\label{eq_algdh}
\begin{align}
    \dot{z}_k^H &=-\!\!\sum_{j\in\left(\mathcal{N}_k\bigcap \mathcal{M}\right)}\!\!(\lambda^H_k-\lambda_j)-
    \!\!\sum_{\ell \in \left(\mathcal{N}_k\bigcap \mathcal{H}\right)}\!\!(\lambda^H_k-\lambda^H_\ell)\\[.5ex]\nonumber\\
    \dot{\lambda}_k^H &= \left[B_k q_k(x_{\mathcal{N}_k}) + \sum_{j\in\left(\mathcal{N}_k\bigcap \mathcal{M}\right)}(z^H_k-z_j)\right.\nonumber\\
    &\qquad\qquad \left.+\sum_{\ell \in \left(\mathcal{N}_k\bigcap \mathcal{H}\right)}(z^H_k-z^H_\ell)+ c^H_k\right]_{{\lambda}_k^H}^+.
\end{align}
\end{subequations}
It can be directly observed that the algorithm (\ref{eq_algda}-\ref{eq_algdh}) can be implemented by using locally available information from the neighbors of each autonomous agent $i$ and human agent $k$, making the algorithm inherently \textbf{distributed}. Furthermore, updates (\ref{eq_algda}-\ref{eq_algdh}) clearly differentiate between the roles of human and autonomous agents in the system. In particular, humans are uncontrollable. Their states $y_k$ do not have an update in \eqref{eq_algdh} but instead depend on human response functions \eqref{eq_response}. In contrast, for the state updates of autonomous agents in \eqref{eq_algda1}, they must account not only for their own optimality, i.e., $\frac{\partial f_i}{\partial x_i}$, and constraints, i.e., $A_i^{\top}\lambda_i$, but also for the optimality, i.e., $\frac{\partial q_\ell}{\partial x_i}\frac{\partial g}{\partial q_\ell} $, and constraints, i.e., $\left(B_\ell \frac{\partial q_\ell}{\partial x_i}\right)^{\top} {\lambda}^H_\ell$, of their neighboring human agents $\ell \in \left(\mathcal{N}_i\bigcap \mathcal{H}\right)$.


\medskip
\begin{theorem}[Effectiveness of the Proposed Algorithm] \label{Thm_1}
Suppose Assumption \ref{Asmp_1} holds. The distributed algorithm defined by update \eqref{eq_alg} or equivalently (\ref{eq_algda}-\ref{eq_algdh}), has the following properties:
\end{theorem}

\begin{enumerate}[label={\alph*}.]
\item $\textit{[Equilibrium]}$ The equilibrium points $(\mathbf{x}^*, \mathbf{z}^*, \boldsymbol{\lambda}^*)$ of \eqref{eq_alg} always exist.  
Any equilibrium point is a saddle point of \eqref{eq_Larg} over the domain $\mathbb{R}^{\sum_{i\in\mathcal{M}}n_i}\times\mathbb{R}^{r(m+h)}\times\mathbb{R}^{r(m+h)}_{\ge0}$. The $\mathbf{x}^*$ gives the global optimal solution to the constrained optimization problem \ref{prob_1};
\item $\textit{[Convergence]}$ Update \eqref{eq_alg} drives the states $\mathbf{x}(t)$, $\mathbf{z}(t)$, and $\boldsymbol{\lambda}(t)$ of all agents to asymptotically converge to an equilibrium point  $(\mathbf{x}^*, \mathbf{z}^*, \boldsymbol{\lambda}^*)$. 
\end{enumerate}

\medskip

The above result for Theorem \ref{Thm_1} is based on the assumption that all human response functions $q_k(\cdot)$ are directly known. However, echoing the Remark \ref{RM_H}, if the algorithm starts by using an inaccurate $\hat{q}_k(\cdot)$, we have the following results.

\begin{corollary}\label{col_1}
Given Assumption 1, we further assume: (i) functions $f_i(\cdot)$, $g_k(\cdot)$, $q_k(\cdot)$ are locally Lipschitz with respect to their arguments; (ii) there exists a time $T<\infty$ such that the approximation of human response function $q_k(\cdot)$ in all virtual proxies converge to their true values for $t>T$. Then the system states in \eqref{eq_alg} is bounded for $t\le T$, and will converge to $(\mathbf{x}^*, \mathbf{z}^*, \boldsymbol{\lambda}^*)$ as $t\to\infty$.
\end{corollary}

\section{Analysis}\label{Sec_ana}

\subsection{Proof of Theorem \ref{Thm_0}}
The coupled constraint \eqref{eq_repfb} can be rewritten as:
\begin{equation}\label{eq_reconst}
(\mathbf{1}_{m+h}^{\top}\otimes I_r)
       \begin{bmatrix} \bar{A} & \\ & \bar{B} \end{bmatrix}
    \begin{bmatrix} \mathbf{x} \\ \mathbf{y} \end{bmatrix}
    + (\mathbf{1}_{m+h}^{\top}\otimes I_r)C \le 0,
\end{equation}
where the notations align with the ones defined in \eqref{eq_lm1}.

Then to complete the proof, we only need to show the equivalence between \eqref{eq_reconst} and \eqref{eq_lm1}. To this end, we first show that \eqref{eq_reconst} implies \eqref{eq_lm1}. By \eqref{eq_reconst}, there must exist a vector $p$, which has the same dimension as $C$, and $p\ge 0$ such that
\begin{align*}
(\mathbf{1}_{m+h}^{\top}\otimes I_r)
       \begin{bmatrix} \bar{A} & \\ & \bar{B} \end{bmatrix}
    \begin{bmatrix} \mathbf{x} \\ \mathbf{y} \end{bmatrix}
    + (\mathbf{1}_{m+h}^{\top}\otimes I_r)(C+p) = 0,
\end{align*}
Consequently,
\begin{align}\label{eq_ABker}
\begin{bmatrix}
        \bar{A}&\\&\bar{B}
    \end{bmatrix}\begin{bmatrix}
        \mathbf{x}\\\mathbf{y}
    \end{bmatrix} +(C + p) \in \ker (\mathbf{1}_{m+h}^{\top}\otimes I_r).
\end{align}


Since $\mathbb{G}$ is connected, $\ker(L)$ is the column span of $\mathbf{1}_{m+h}$. It follows that
$L\mathbf{1}_{m+h}=0$ and $\mathbf{1}_{m+h}^{\top}L^{\top}=0.$
Thus, $\ker(\mathbf{1}_{m+h}^{\top})=\operatorname{image}(L)$.
Since $\mathbb{G}$ is undirected, $L$ is symmetric. Using the properties of Kronecker product, one has  
\begin{align}
\ker(\mathbf{1}_{m+h}^{\top}\otimes I_r) = \operatorname{image}(L\otimes I_r)=\operatorname{image}(\bar{L}).
\end{align}
This and \eqref{eq_ABker} imply 
\begin{align}
\begin{bmatrix}
        \bar{A}&\\&\bar{B}
    \end{bmatrix}\begin{bmatrix}
        \mathbf{x}\\\mathbf{y}
    \end{bmatrix} +(C + p) \in \operatorname{image}(\bar{L}).
\end{align}
Thus there always exists a $\mathbf{z}$ such that 
\begin{align}
\begin{bmatrix}
        \bar{A}&\\&\bar{B}
    \end{bmatrix}\begin{bmatrix}
        \mathbf{x}\\ \mathbf{y}
    \end{bmatrix} +(C + p) +\bar{L}\mathbf{z}=0
\end{align}
Since $p\ge 0$, one has \eqref{eq_lm1} is true.

Second, we prove \eqref{eq_lm1} implies \eqref{eq_reconst}. By \eqref{eq_lm1}, one has
\begin{align}
(\mathbf{1}_{m+h}^{\top}\otimes I_r)\left(\begin{bmatrix}
        \bar{A}&\\&\bar{B}
    \end{bmatrix}\begin{bmatrix}
        \mathbf{x}\\ \mathbf{y}
    \end{bmatrix} +\bar L\mathbf{z}+C\right)\le0,\nonumber
\end{align}
from which and the fact that $(\mathbf{1}_{m+h}^{\top}\otimes I_r)\bar L= (\mathbf{1}_{m+h}^{\top}L)\otimes I_r = 0$ for undirected and connected $\mathbb{G}$, one has \eqref{eq_reconst}. 
\hfill \qed




\smallskip

\subsection{Proof of Theorem \ref{Thm_1}}
\subsubsection{Statement a}
We first show that the equilibrium points of \eqref{eq_alg} exist.
From Assumption \ref{Asmp_1}, the objective and constraint functions in (\ref{eq_repf}a-b) are convex. Furthermore, recall our assumption that there exists $x_i\in \mathbb{R}^{n_i}$ such that $\sum_{i\in\mathcal{M}} A_i x_i + \sum_{k\in\mathcal{H}} B_k q_k(x_{\mathcal{N}_{k}}) + c < 0$. Thus, problem \eqref{eq_repf} meets the Slater condition~\cite{slater2013lagrange}.
Consequently, strong duality holds and there exists points $(\mathbf{x}^*, \mathbf{z}^*, \boldsymbol{\lambda}^*)$ satisfying the following Karush-Kuhn-Tucker (KKT) conditions~\cite[Sec. 5.5.3]{boyd2004convex},
{\small\begin{subequations}\label{eq_KKT}
    \begin{align}\label{eq_KKTa}
    \left.\frac{\partial\mathcal{L}(\mathbf{x},\mathbf{z},\boldsymbol{\lambda})}{\partial \mathbf{x}}\right|_{\substack{
    \mathbf{x}=\mathbf{x}^*\\ \boldsymbol{\lambda}=\boldsymbol{\lambda}^*
    }}=0,\quad \left.\frac{\partial\mathcal{L}(\mathbf{x},\mathbf{z},\boldsymbol{\lambda})}{\partial \mathbf{z}}\right|_{\boldsymbol{\lambda}=\boldsymbol{\lambda}^*}=0, \\\label{eq_KKTb}
    \begin{bmatrix}\bar{A}\mathbf{x}^*\\\bar{B}Q(\mathbf{x}^*)\end{bmatrix}+\bar L\mathbf{z}^*+C\le0,\qquad \bm{\lambda}^*\ge0\\  \label{eq_KKTc}  \boldsymbol{{\lambda^*}}^{\top}\left(\begin{bmatrix}\bar{A}\mathbf{x}^*\\\bar{B}Q(\mathbf{x}^*)\end{bmatrix}+\bar L\mathbf{z}^*+C\right)=0
\end{align}
\end{subequations}
}Clearly, condition \eqref{eq_KKTa} implies the equilibrium for (\ref{eq_alg}a-b); while conditions (\ref{eq_KKT}b-c) imply the equilibrium for (\ref{eq_alg}c).

Furthermore, since the problem satisfies the Slater condition, the points $(\mathbf{x}^*, \mathbf{z}^*, \boldsymbol{\lambda}^*)$ must be a saddle point of \eqref{eq_Larg} over the domain $\mathbb{R}^{\sum_{i\in\mathcal{M}}n_i}\times\mathbb{R}^{r(m+h)}\times\mathbb{R}^{r(m+h)}_{\ge0}$~\cite[Sec. 3]{cherukuri2016asymptotic}. Furthermore, due to KKT condition~\eqref{eq_KKT}, the equilibrium $\mathbf{x}^*$ gives the global optimal solution to the constrained optimization problem.

\subsubsection{Statement b} The proof of convergence appears commonly in distributed optimization, which follows mainly from existing results in~\cite{cherukuri2016asymptotic}. We do not claim it as our main contribution. The proof is given in the appendix for completeness.
\hfill \qed

\subsection{Proof of Corollary  \ref{col_1}}
Given the local Lipschitz continuity of $f_i(\cdot)$, $g_k(\cdot)$, $q_k(\cdot)$, the terms $\frac{\partial F}{\partial\mathbf{x}}, \frac{\partial G}{\partial{Q}}, \frac{\partial Q}{\partial\mathbf{x}}$ in \eqref{eq_alg} are bounded if the system states are bounded. Since $T$ is finite, both system states and $\frac{\partial F}{\partial\mathbf{x}}, \frac{\partial G}{\partial{Q}}, \frac{\partial Q}{\partial\mathbf{x}}$ must be bounded.


For the second statement, the convergence follows directly from Theorem \ref{Thm_1}, assuming that the states are initialized using $(\mathbf{x}(t=T), \mathbf{z}(t=T), \boldsymbol{\lambda}(t=T))$.
\hfill \qed

\section{Simulated Experiments}
We consider a system involving five autonomous agents and two human workers collaboratively executing production tasks. Each autonomous agent and human has a specific production capability for manufacturing different types of parts. The objective is to minimize the production cost using the proposed distributed algorithm, while adhering to resource constraints and productivity requirements.

\textit{Simulation Setup and System Description:}
The state vector $x_i$ of each autonomous agent $i$ represents the quantities of $n_i$ types of parts it produces. The number of part types handled by each autonomous agent is: $n_1 = 3$, $n_2 = 5$, $n_3 = 4$, $n_4 = 2$,  $n_5 = 1$. Similarly, the state vector $y_k$ of human $k$ represents the quantities of $m_k$ types of parts they supervise or assist with. The number of part types for each human is $m_1 = 3$ and $m_2 = 5$.
The state of each autonomous agent and human worker is influenced by its neighboring agents, as indicated in Fig. 1. 
The network is undirected and connected.

The production cost of each autonomous agent and human is represented by the following quadratic functions:
\begin{align}
f_i(x_i) &= x_i^\top \Lambda_i x_i, \nonumber \\
g_k(y_k) &= y_k^\top \Gamma_k y_k,
\end{align}
where $\Lambda_i$ and $\Gamma_k$ are positive definite matrices, which map the types and quantities of parts produced by the autonomous agents and humans to their costs.


According to Sec. \ref{Sec_HM}, the response function of each human is modeled by the combination of the subjective utility functions and a neural network:
\begin{align*}
y_k = q_k(x_{\mathcal{N}_k};\theta_k)
\end{align*}
Let $\theta_k$ be the learned parameters of the human model. We assume that humans are more reliable in conducting tasks but can get fatigue, and are more expensive, while autonomous agents are less expensive and suitable for dull tasks but are less reliable. The cost of task failure will be much more expensive than the human labor cost. Correspondingly, we introduce two sets of parameters representing: (i) risk-seeking humans, who are more comfortable in assigning more tasks to autonomous agents (smaller $y_k$) after observing other agents' states $x_{\mathcal{N}k}$; (ii) risk-averse humans, who prefer to work more by themselves (larger $y_k$) after observing other agents' states $x_{\mathcal{N}_k}$.

The formulation of \eqref{eq_pfb} set up several globally coupled constraints. 
Specifically, we consider resource constraints where the total available working hours (and energy) for the entire production process are embedded in $c$. The corresponding entries in $A_i$ and $B_k$ map the production of different parts $x_i$ and $y_k$ to the required working hours (energy consumption) by each agent. 
We also consider the production output requirement, where $A_i$ and $B_k$ map how the produced parts can be assembled into final products, and $c$ represents the total demand for final products. 

\textit{Simulation Results:}
To demonstrate the convergence of our distributed resource allocation algorithm, one human agent is risk-seeking and one human agent is risk-averse.
We run the proposed algorithm and consider the following function:
\[
W(t) = \sum_{i=1}^5 \|x_i(t) - x_i^*\|_2^2 +\sum_{k=1}^2 \|y_k(t) - y_k^*\|_2^2,
\]
which measures the deviation of the current production state $x_i(t)$ from the optimal solution $x_i^*,y^*$ that we manually compute. Since $\Lambda_i$ and $\Gamma_k$ are positive definite, the optimal solution is unique.  As shown in Fig. \ ref {fig: simulation}, $W(t)$ converges to zero, indicating that system states approach the optimal production state over time. This simulation agrees with our main result in Theorem \ref{Thm_1}.


\begin{figure}[t]
    \centering
    \includegraphics[width=0.8\linewidth]{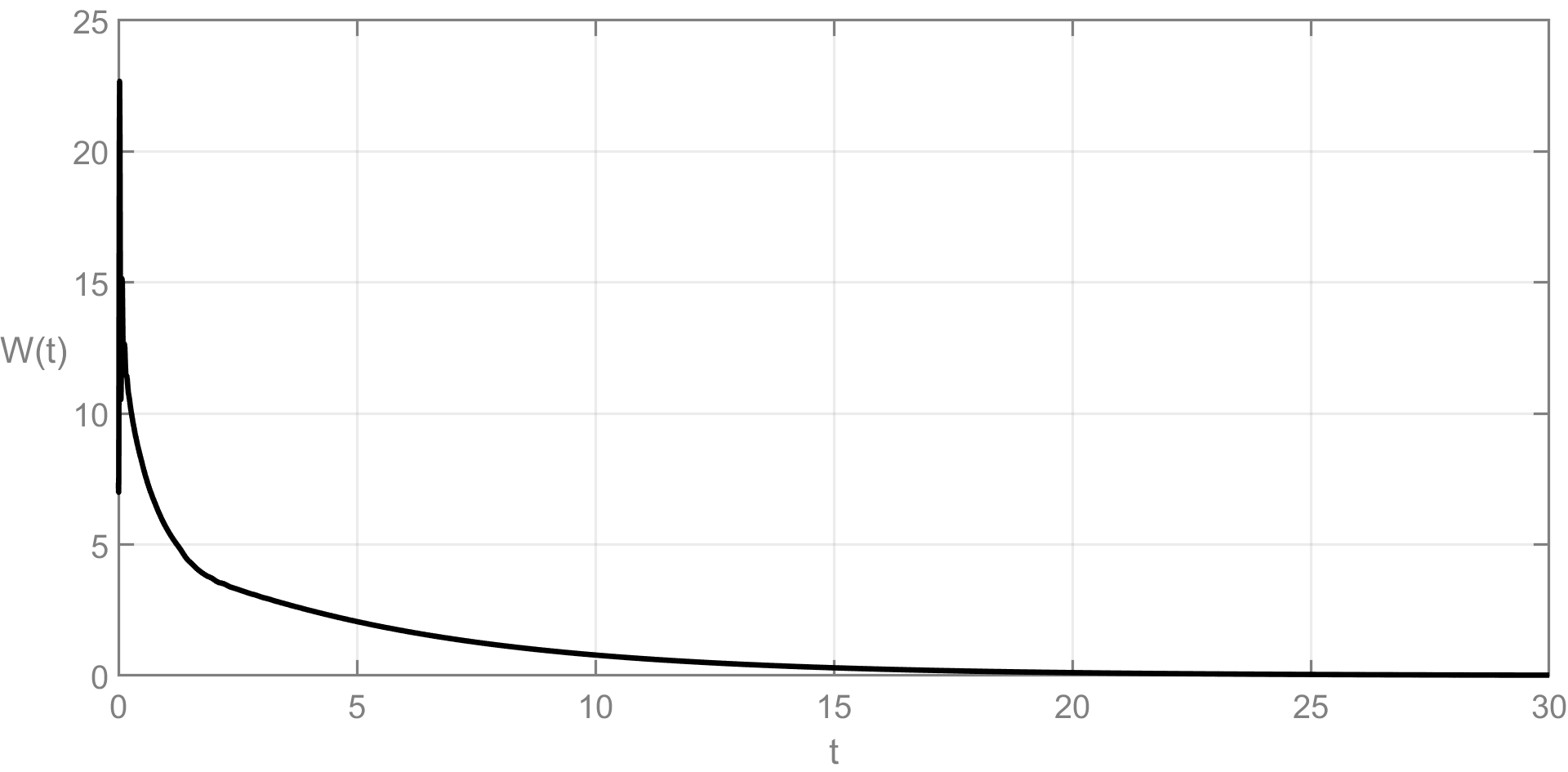}
    \caption{Convergence indicated by evolution of $W(t)$}
    \label{fig:simulation}
    \vspace{-1em}
\end{figure}


We also examine the distribution of work between humans and autonomous agents when the involved humans have different risk attitudes. This will demonstrate the capability of our algorithm, where autonomous agents will adapt to human preferences when optimizing overall system performance. In our setup, human 1 is in general more efficient than human 2 in terms of cost $g_k$, i.e., $\|\Gamma_1\|_2<\|\Gamma_2\|_2$. We use the one norm $\|\cdot\|_1$ of the states to represent the workload of agents.

\begin{figure}[t]
    \centering
    \includegraphics[width=0.45\textwidth]{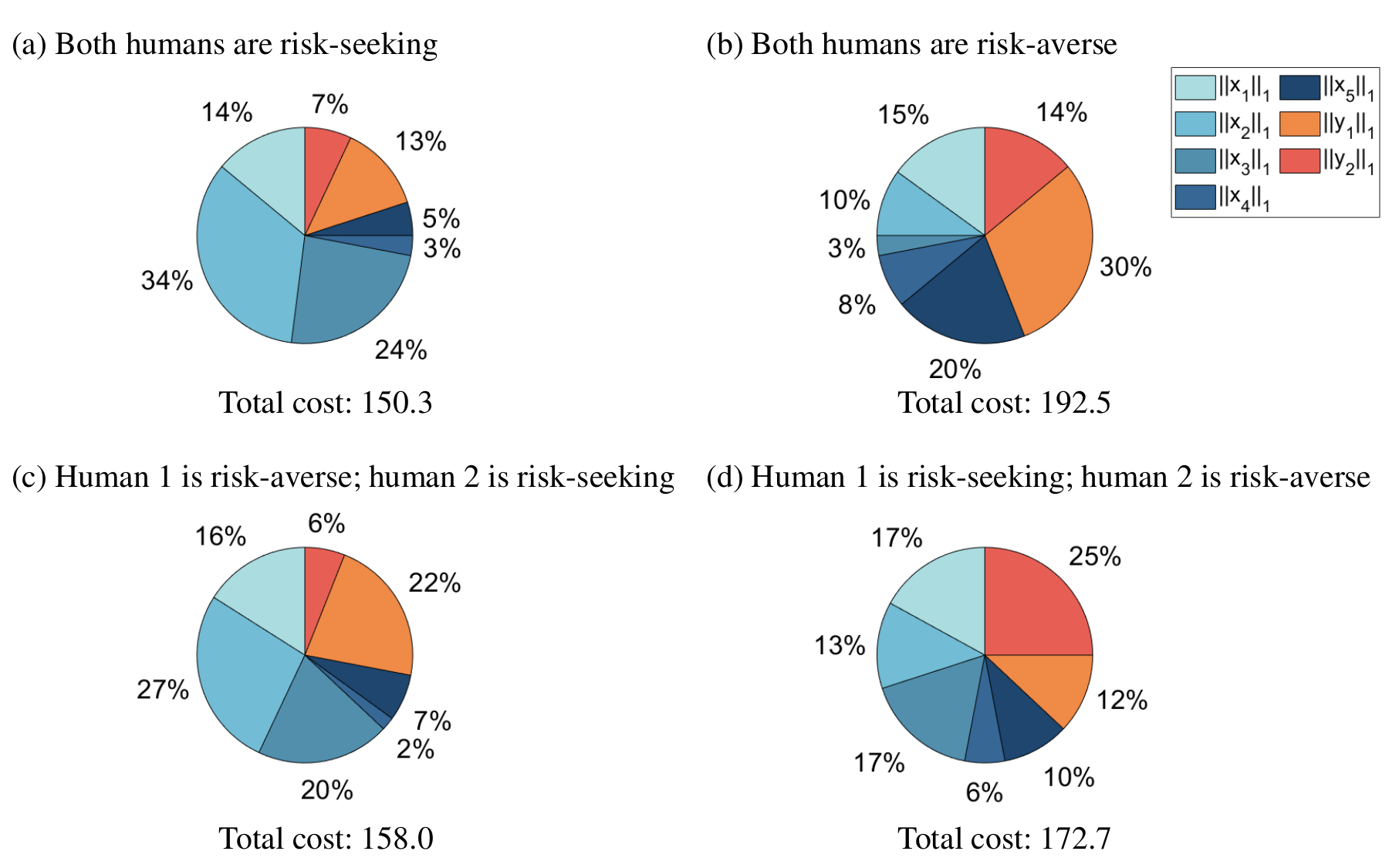}
    \caption{Proportion of work consumed by humans versus autonomous agents and the system cost. }
    \label{fig:scenario1}
    \vspace{-1.5em}
\end{figure}

As shown in Fig.~\ref{fig:scenario1}, when risk-seeking human agents are involved, the autonomous agents increase their workload to ensure all constraints are satisfied, and the opposite occurs when more humans are risk-averse. Furthermore, we observe that the autonomous agents adapt their states to encourage human 1 to work more (based on their response function), as human 1 is more efficient and helps reduce overall costs. Finally, by comparing the overall system costs in cases (c) and (d), when human 1 is risk-averse and takes on more tasks, the overall system cost is smaller.

\section{Conclusion}
We have considered distributed resource allocation for human-autonomy teaming under coupled constraints. We have modeled human behaviors using a response function that captures human biases and risk attitudes during decision-making processes. The analysis of the coupled constraints between autonomous agents and humans has led us to introduce an equivalent reformulation of the constraint, which is decoupled over the network topology of the system. Building on this, we have proposed a distributed algorithm that enables autonomous agents to cooperatively update their states while accounting for human responses. Our theoretical analysis has demonstrated that the combined states of the system and human responses achieve globally optimal resource allocation. We have illustrated the effectiveness of the proposed algorithm in simulation. Future work will extend the current undirected network topology to directed ones and consider more complex dynamics in human modeling.

\section*{Appendix}

\noindent\textbf{{{Proof of Theorem \ref{Thm_1}-b.}}}
To facilitate the subsequent analysis, we introduce a composite vector $\bm{\eta}=\begin{bmatrix}\mathbf{x} & \mathbf{z}\end{bmatrix}^{\top}\in\mathbb{R}^{\sum_{i\in\mathcal{M}}n_i+r(m+h)}$, and $\mathcal{L}(\bm{\eta}, \boldsymbol{\lambda})$ retains its convexity in $\bm{\eta}$ and linearity in $\boldsymbol{\lambda}$ due to Assumption~\ref{Asmp_1}. We denote a saddle point of \eqref{eq_Larg} over the domain $\mathbb{R}^{\sum_{i\in\mathcal{M}}n_i+r(m+h)}\times\mathbb{R}^{r(m+h)}_{\ge0}$ as $\bm{\eta}^\dagger=\begin{bmatrix}\mathbf{x}^\dagger & \mathbf{z}^\dagger\end{bmatrix}^{\top}$ and $\boldsymbol{\lambda}^\dagger$.
Consider a positive definite scalar-valued function $V$ of the following form:
\begin{align}
V(\bm{\eta},  \boldsymbol{\lambda}) = \frac{1}{2} \|\bm{\eta} - \bm{\eta}^\dagger\|^2 + \frac{1}{2} \|\boldsymbol{\lambda} - \boldsymbol{\lambda}^\dagger\|^2
\end{align}

The derivative of \(V(\bm{\eta}, \boldsymbol{\lambda})\) along dynamics \eqref{eq_alg} can be computed as:
{\small\begin{align}
\dot{V}(\bm{\eta}, \boldsymbol{\lambda}) &= \frac{\partial V}{\partial \bm{\eta}} \dot{\bm{\eta}} + \frac{\partial V}{\partial \boldsymbol{\lambda}} \dot{\boldsymbol{\lambda}} \nonumber \\
&=\!-(\bm{\eta} \!-\! \bm{\eta}^\dagger)^\top \frac{\partial \mathcal{L}(\bm{\eta},  \boldsymbol{\lambda})}{\partial \bm{\eta}} \nonumber \!+\! (\boldsymbol{\lambda} \!-\! \boldsymbol{\lambda}^\dagger)^\top \left[\frac{\partial \mathcal{L}(\bm{\eta}, \boldsymbol{\lambda})}{\partial \boldsymbol{\lambda}}\right]_{\boldsymbol{\lambda}}^{+} \nonumber \\
& = -(\bm{\eta} - \bm{\eta}^\dagger)^\top \frac{\partial \mathcal{L}(\bm{\eta},  \boldsymbol{\lambda})}{\partial \bm{\eta}} + (\boldsymbol{\lambda} - \boldsymbol{\lambda}^*)^\top \frac{\partial \mathcal{L}(\bm{\eta}, \boldsymbol{\lambda})}{\partial \boldsymbol{\lambda}}\nonumber\\
&\quad +(\boldsymbol{\lambda} - \boldsymbol{\lambda}^*)^\top \left(\left[\frac{\partial \mathcal{L}(\bm{\eta}, \boldsymbol{\lambda})}{\partial \boldsymbol{\lambda}}\right]_{\boldsymbol{\lambda}}^{+} -\frac{\partial \mathcal{L}(\bm{\eta}, \boldsymbol{\lambda})}{\partial \boldsymbol{\lambda}} \right).\nonumber
\end{align}
}Since $\mathcal{L}$ is convex in $\bm{\eta}$ and concave in $\bm{\lambda}$, for all $(\bm{\eta},\bm{\lambda}) \in\mathbb{R}^{\sum_{i\in\mathcal{M}}n_i+r(m+h)}\times\mathbb{R}^{r(m+h)}_{\ge0}$, the corresponding first-order conditions yield the following bound:
{\small\begin{align}
\dot{V}(\bm{\eta}, \boldsymbol{\lambda}) & \leq \mathcal{L}(\bm{\eta}^\dagger, \boldsymbol{\lambda}) - \mathcal{L}(\bm{\eta},  \boldsymbol{\lambda}) + \left(\mathcal{L}(\bm{\eta}, \boldsymbol{\lambda}) - \mathcal{L}(\bm{\eta},  \boldsymbol{\lambda}^\dagger) \right) \nonumber \\
&\quad +(\boldsymbol{\lambda} - \boldsymbol{\lambda}^\dagger)^\top \left(\left[\frac{\partial \mathcal{L}(\bm{\eta}, \boldsymbol{\lambda})}{\partial \boldsymbol{\lambda}}\right]_{\boldsymbol{\lambda}}^{+} -\frac{\partial \mathcal{L}(\bm{\eta}, \boldsymbol{\lambda})}{\partial \boldsymbol{\lambda}} \right)\nonumber\\
&= \mathcal{L}(\bm{\eta}^\dagger, \boldsymbol{\lambda}) - \mathcal{L}(\bm{\eta}^\dagger, \boldsymbol{\lambda}^\dagger) \nonumber +  \mathcal{L}(\bm{\eta}^\dagger,  \boldsymbol{\lambda}^\dagger) - \mathcal{L}(\bm{\eta}, \boldsymbol{\lambda}^\dagger)\\
&\quad +(\boldsymbol{\lambda} - \boldsymbol{\lambda}^\dagger)^\top \left(\left[\frac{\partial \mathcal{L}(\bm{\eta}, \boldsymbol{\lambda})}{\partial \boldsymbol{\lambda}}\right]_{\boldsymbol{\lambda}}^{+} -\frac{\partial \mathcal{L}(\bm{\eta}, \boldsymbol{\lambda})}{\partial \boldsymbol{\lambda}} \right)\nonumber\\
& \leq 0 \label{eq_Vdot}
\end{align}
}where the first two terms of the last inequality arise from the definition of the saddle point in Theorem \ref{Thm_1}-a. To analyze the last term, let $(\cdot)_\ell$ denote the $\ell$'th entry of a vector. For any entry of $\boldsymbol{\lambda}$, if $(\boldsymbol{\lambda})_\ell >0$, then $\left[\left(\frac{\partial \mathcal{L}(\bm{\eta}, \boldsymbol{\lambda})}{\partial \boldsymbol{\lambda}}\right)_{\ell}\right]_{\boldsymbol{(\lambda})_{\ell}}^{+} -\left(\frac{\partial \mathcal{L}(\bm{\eta}, \boldsymbol{\lambda})}{\partial \boldsymbol{\lambda}}\right)_{\ell} = 0 $. If $(\boldsymbol{\lambda})_\ell=0$, then $((\boldsymbol{\lambda})_\ell - (\boldsymbol{\lambda})_\ell^\dagger) \leq 0$ and $\left[\left(\frac{\partial \mathcal{L}(\bm{\eta}, \boldsymbol{\lambda})}{\partial \boldsymbol{\lambda}}\right)_{\ell}\right]_{(\boldsymbol{\lambda})_{\ell}}^{+} -\left(\frac{\partial \mathcal{L}(\bm{\eta}, \boldsymbol{\lambda})}{\partial \left(\boldsymbol{\lambda}\right)_{\ell}}\right)_{\ell} \geq 0 $, 
which implies that $((\boldsymbol{\lambda})_\ell - (\boldsymbol{\lambda})_\ell^\dagger)^{\top}\left(\left[\left(\frac{\partial \mathcal{L}(\bm{\eta}, \boldsymbol{\lambda})}{\partial \boldsymbol{\lambda}}\right)_{\ell}\right]_{(\boldsymbol{\lambda})_{\ell}}^{+} -\left(\frac{\partial \mathcal{L}(\bm{\eta}, \boldsymbol{\lambda})}{\partial\boldsymbol{\lambda}}\right)_{\ell}\right)\le0
$. Combining these independent entries together yields
$(\boldsymbol{\lambda} - \boldsymbol{\lambda}^*)^\top \left(\left[\frac{\partial \mathcal{L}(\bm{\eta}, \boldsymbol{\lambda})}{\partial \boldsymbol{\lambda}}\right]_{\boldsymbol{\lambda}}^{+} -\frac{\partial \mathcal{L}(\bm{\eta}, \boldsymbol{\lambda})}{\partial \boldsymbol{\lambda}} \right) \leq 0$. Consequently. $\dot{V}(\bm{\eta}, \boldsymbol{\lambda}) \leq 0$.

Based on \eqref{eq_Vdot} and the LaSalle's invariance principle, the trajectory of \eqref{eq_alg} converges asymptotically to the invariant set $\mathcal{S}=\{({\bm\eta}^*,{\bm\lambda}^*)~|~\dot{V} = 0\}$. $\forall ({\bm\eta}^*,{\bm\lambda}^*) \in \mathcal{S}$, and there holds
\begin{subequations}\label{eq_ieq}
    \begin{align}
    \mathcal{L}(\bm{\eta}^\dagger, \boldsymbol{\lambda}^*) - \mathcal{L}(\bm{\eta}^\dagger, \boldsymbol{\lambda}^\dagger)&=0\label{eq_ieqa}\\
    \mathcal{L}(\bm{\eta}^\dagger,  \boldsymbol{\lambda}^\dagger) - \mathcal{L}(\bm{\eta}^*, \boldsymbol{\lambda}^\dagger)&=0\label{eq_ieqb}\\
    (\boldsymbol{\lambda}^* - \boldsymbol{\lambda}^\dagger)^\top \left(\left[\frac{\partial \mathcal{L}(\bm{\eta}, \boldsymbol{\lambda})}{\partial \boldsymbol{\lambda}}\right]_{\boldsymbol{\lambda}}^{+} -\frac{\partial \mathcal{L}(\bm{\eta}, \boldsymbol{\lambda})}{\partial \boldsymbol{\lambda}} \right)&=0\label{eq_ieqc},
    \end{align}
\end{subequations}
where in \eqref{eq_ieqc}, the partial derivatives are evaluated at $({\bm\eta}^*,{\bm\lambda}^*)$.
Since $(\bm{\eta}^\dagger,  \boldsymbol{\lambda}^\dagger)$ is a saddle point of $\mathcal{L}(\cdot)$ over the domain $\mathbb{R}^{\sum_{i\in\mathcal{M}}n_i+r(m+h)}\times\mathbb{R}^{r(m+h)}_{\ge0}$, \eqref{eq_ieq} implies that any $({\bm\eta}^*,{\bm\lambda}^*) \in \mathcal{S}$ is also a saddle point of \eqref{eq_Larg} over the same domain~\cite{cherukuri2016asymptotic}. Then by Theorem \ref{Thm_1}-a, the corresponding entries of $x_i$ in all agents, together with human response $y_k$, solve the original constrained optimization problem \eqref{eq_pf}, satisfying all local constraints and achieving global optimality.
\hfill \qed



\bibliographystyle{IEEEtran}
\bibliography{references}

\begin{thebibliography}{10}
\providecommand{\url}[1]{#1}
\csname url@samestyle\endcsname
\providecommand{\newblock}{\relax}
\providecommand{\bibinfo}[2]{#2}
\providecommand{\BIBentrySTDinterwordspacing}{\spaceskip=0pt\relax}
\providecommand{\BIBentryALTinterwordstretchfactor}{4}
\providecommand{\BIBentryALTinterwordspacing}{\spaceskip=\fontdimen2\font plus
\BIBentryALTinterwordstretchfactor\fontdimen3\font minus \fontdimen4\font\relax}
\providecommand{\BIBforeignlanguage}[2]{{%
\expandafter\ifx\csname l@#1\endcsname\relax
\typeout{** WARNING: IEEEtran.bst: No hyphenation pattern has been}%
\typeout{** loaded for the language `#1'. Using the pattern for}%
\typeout{** the default language instead.}%
\else
\language=\csname l@#1\endcsname
\fi
#2}}
\providecommand{\BIBdecl}{\relax}
\BIBdecl

\bibitem{cao2021human}
M.~Cao, ``Human decision-making in multi-agent systems,'' in \emph{Encyclopedia of Systems and Control}.\hskip 1em plus 0.5em minus 0.4em\relax Springer, 2021, pp. 909--913.

\bibitem{kwon2020humans}
M.~Kwon, E.~Biyik, A.~Talati, K.~Bhasin, D.~P. Losey, and D.~Sadigh, ``When humans aren't optimal: Robots that collaborate with risk-aware humans,'' in \emph{Proceedings of the 2020 ACM/IEEE international conference on human-robot interaction}, 2020, pp. 43--52.

\bibitem{wang2018distributed}
X.~Wang and S.~Mou, ``A distributed algorithm for achieving the conservation principle,'' in \emph{2018 Annual American Control Conference (ACC)}.\hskip 1em plus 0.5em minus 0.4em\relax IEEE, 2018, pp. 5863--5867.

\bibitem{zhu2011distributed}
M.~Zhu and S.~Martinez, ``On distributed convex optimization under inequality and equality constraints,'' \emph{IEEE Transactions on Automatic Control}, vol.~57, no.~1, pp. 151--164, 2011.

\bibitem{zulhasnine2010efficient}
M.~Zulhasnine, C.~Huang, and A.~Srinivasan, ``Efficient resource allocation for device-to-device communication underlaying lte network,'' in \emph{2010 IEEE 6th International conference on wireless and mobile computing, networking and communications}.\hskip 1em plus 0.5em minus 0.4em\relax IEEE, 2010, pp. 368--375.

\bibitem{tang2015resource}
J.~Tang, D.~K. So, E.~Alsusa, K.~A. Hamdi, and A.~Shojaeifard, ``Resource allocation for energy efficiency optimization in heterogeneous networks,'' \emph{IEEE Journal on Selected Areas in Communications}, vol.~33, no.~10, pp. 2104--2117, 2015.

\bibitem{morariu2020machine}
C.~Morariu, O.~Morariu, S.~R{\u{a}}ileanu, and T.~Borangiu, ``Machine learning for predictive scheduling and resource allocation in large scale manufacturing systems,'' \emph{Computers in Industry}, vol. 120, p. 103244, 2020.

\bibitem{xiao2004simultaneous}
L.~Xiao, M.~Johansson, and S.~P. Boyd, ``Simultaneous routing and resource allocation via dual decomposition,'' \emph{IEEE Transactions on Communications}, vol.~52, no.~7, pp. 1136--1144, 2004.

\bibitem{shorinwa2023distributed}
O.~Shorinwa, R.~N. Haksar, P.~Washington, and M.~Schwager, ``Distributed multirobot task assignment via consensus admm,'' \emph{IEEE Transactions on Robotics}, vol.~39, no.~3, pp. 1781--1800, 2023.

\bibitem{jiang2021distributed}
W.~Jiang and T.~Charalambous, ``Distributed alternating direction method of multipliers using finite-time exact ratio consensus in digraphs,'' in \emph{2021 European Control Conference (ECC)}.\hskip 1em plus 0.5em minus 0.4em\relax IEEE, 2021, pp. 2205--2212.

\bibitem{falsone2020tracking}
A.~Falsone, I.~Notarnicola, G.~Notarstefano, and M.~Prandini, ``Tracking-admm for distributed constraint-coupled optimization,'' \emph{Automatica}, vol. 117, p. 108962, 2020.

\bibitem{banjac2019decentralized}
G.~Banjac, F.~Rey, P.~Goulart, and J.~Lygeros, ``Decentralized resource allocation via dual consensus admm,'' in \emph{2019 American Control Conference (ACC)}.\hskip 1em plus 0.5em minus 0.4em\relax IEEE, 2019, pp. 2789--2794.

\bibitem{nedic2018improved}
A.~Nedi{\'c}, A.~Olshevsky, and W.~Shi, ``Improved convergence rates for distributed resource allocation,'' in \emph{2018 IEEE Conference on Decision and Control (CDC)}.\hskip 1em plus 0.5em minus 0.4em\relax IEEE, 2018, pp. 172--177.

\bibitem{aybat2019distributed}
N.~S. Aybat and E.~Y. Hamedani, ``A distributed admm-like method for resource sharing over time-varying networks,'' \emph{SIAM Journal on Optimization}, vol.~29, no.~4, pp. 3036--3068, 2019.

\bibitem{zhu2019distributed}
Y.~Zhu, W.~Ren, W.~Yu, and G.~Wen, ``Distributed resource allocation over directed graphs via continuous-time algorithms,'' \emph{IEEE Transactions on Systems, Man, and Cybernetics: Systems}, vol.~51, no.~2, pp. 1097--1106, 2019.

\bibitem{doan2017distributed}
T.~T. Doan and C.~L. Beck, ``Distributed lagrangian methods for network resource allocation,'' in \emph{2017 IEEE Conference on Control Technology and Applications (CCTA)}.\hskip 1em plus 0.5em minus 0.4em\relax IEEE, 2017, pp. 650--655.

\bibitem{tverskyJudgmentUncertaintyHeuristics1974}
A.~Tversky and D.~Kahneman, ``Judgment under {{Uncertainty}}: {{Heuristics}} and {{Biases}}: {{Biases}} in judgments reveal some heuristics of thinking under uncertainty.'' \emph{Science}, vol. 185, no. 4157, pp. 1124--1131, 1974.

\bibitem{kahnemanThinkingFastSlow2011}
D.~Kahneman, \emph{Thinking, Fast and Slow}, 1st~ed.\hskip 1em plus 0.5em minus 0.4em\relax {Farrar, Straus and Giroux}, 2011.

\bibitem{machinaEncyclopediaActuarialScience2004}
\BIBentryALTinterwordspacing
M.~J. Machina, \emph{Encyclopedia of Actuarial Science}, J.~L. Teugels and B.~Sundt, Eds.\hskip 1em plus 0.5em minus 0.4em\relax Wiley. [Online]. Available: \url{https://econweb.ucsd.edu/~mmachina/papers/Machina_Encyc_of_Actuarial_Science.pdf}
\BIBentrySTDinterwordspacing

\bibitem{kobberlingPreferenceFoundationsNonexpected2003}
V.~Köbberling and P.~P. Wakker, ``Preference {{Foundations}} for {{Nonexpected Utility}}: {{A Generalized}} and {{Simplified Technique}},'' \emph{Mathematics of Operations Research}, vol.~28, no.~3, pp. 395--423, 2003.

\bibitem{guan_cdc_2019}
Y.~Guan, A.~M. Annaswamy, and H.~E. Tseng, ``Cumulative prospect theory based dynamic pricing for shared mobility on demand services,'' \emph{IEEE 58th Conference on Decision and Control (CDC)}, pp. 2239--2244, 2019.

\bibitem{jiang2022risk}
L.~Jiang and Y.~Wang, ``Risk-aware decision-making in human-multi-robot collaborative search: a regret theory approach,'' \emph{Journal of Intelligent \& Robotic Systems}, vol. 105, no.~2, p.~40, 2022.

\bibitem{jiang2022risk-TITS}
L.~Jiang, D.~Chen, Z.~Li, and Y.~Wang, ``Risk representation, perception, and propensity in an integrated human lane-change decision model,'' \emph{IEEE Transactions on Intelligent Transportation Systems}, vol.~23, no.~12, pp. 23\,474--23\,487, 2022.

\bibitem{roeMultialternativeDecisionField2001a}
R.~M. Roe, J.~R. Busemeyer, and J.~T. Townsend, ``Multialternative decision field theory: {{A}} dynamic connectionst model of decision making,'' \emph{Psychological Review}, vol. 108, no.~2, pp. 370--392, 2001.

\bibitem{hamalainen2023differentiable}
A.~H{\"a}m{\"a}l{\"a}inen, M.~M. {\c{C}}elikok, and S.~Kaski, ``Differentiable user models,'' in \emph{Uncertainty in Artificial Intelligence}.\hskip 1em plus 0.5em minus 0.4em\relax PMLR, 2023, pp. 798--808.

\bibitem{slater2013lagrange}
M.~Slater, ``Lagrange multipliers revisited,'' in \emph{Traces and emergence of nonlinear programming}.\hskip 1em plus 0.5em minus 0.4em\relax Springer, 2013, pp. 293--306.

\bibitem{sadrfaridpour2017collaborative}
B.~Sadrfaridpour and Y.~Wang, ``Collaborative assembly in hybrid manufacturing cells: An integrated framework for human--robot interaction,'' \emph{IEEE Transactions on Automation Science and Engineering}, vol.~15, no.~3, pp. 1178--1192, 2017.

\bibitem{kahnemanProspectTheoryAnalysis1979}
D.~Kahneman and A.~Tversky, ``Prospect {{Theory}}: {{An Analysis}} of {{Decision}} under {{Risk}},'' \emph{Econometrica}, vol.~47, no.~2, p. 263, 1979.

\bibitem{bleichrodt2010quantitative}
H.~Bleichrodt, A.~Cillo, and E.~Diecidue, ``A quantitative measurement of regret theory,'' \emph{Management Science}, vol.~56, no.~1, pp. 161--175, 2010.

\bibitem{busemeyerDecisionFieldTheory1993}
J.~R. Busemeyer and J.~T. Townsend, ``Decision field theory: {A} dynamic-cognitive approach to decision making in an uncertain environment,'' \emph{Psychological Review}, vol. 100, no.~3, pp. 432--459, 1993.

\bibitem{quiggin1994regret}
J.~Quiggin, ``Regret theory with general choice sets,'' \emph{Journal of Risk and Uncertainty}, vol.~8, pp. 153--165, 1994.

\bibitem{owen2013game}
G.~Owen, \emph{Game theory}.\hskip 1em plus 0.5em minus 0.4em\relax Emerald Group Publishing, 2013.

\bibitem{nedic2009distributed}
A.~Nedic and A.~Ozdaglar, ``Distributed subgradient methods for multi-agent optimization,'' \emph{IEEE Transactions on Automatic Control}, vol.~54, no.~1, pp. 48--61, 2009.

\bibitem{bertsekasnonlinear}
D.~P. Bertsekas and A.~Scientific, ``Nonlinear programming 2nd edition solutions manual.''

\bibitem{boyd2004convex}
S.~Boyd and L.~Vandenberghe, \emph{Convex optimization}.\hskip 1em plus 0.5em minus 0.4em\relax Cambridge university press, 2004.

\bibitem{cherukuri2016asymptotic}
A.~Cherukuri, E.~Mallada, and J.~Cort{\'e}s, ``Asymptotic convergence of constrained primal--dual dynamics,'' \emph{Systems \& Control Letters}, vol.~87, pp. 10--15, 2016.

\end{thebibliography}

    
\end{document}